# SIMUS: an open-source simulator for medical ultrasound imaging.
## Part I: theory & examples

Damien Garcia

*Background and Objective* – Computational ultrasound imaging has become a well-established methodology in the ultrasound community. Simulations of ultrasound sequences and images allow the study of innovative techniques in terms of emission strategy, beamforming, and probe design. There is a wide spectrum of software dedicated to ultrasound imaging, each having its specificities in its applications and the numerical method.

*Methods* – We describe in this two-part paper a new ultrasound simulator (SIMUS) for MATLAB, which belongs to the MATLAB UltraSound Toolbox (MUST). The SIMUS software simulates acoustic pressure fields and radiofrequency RF signals for uniform linear or convex probes. SIMUS is an open-source software whose features are 1) ease of use, 2) time-harmonic analysis, 3) pedagogy. The main goal was to offer a comprehensive turnkey tool, along with a detailed theory for pedagogical and research purposes.

*Results* – This article describes in detail the underlying linear theory of SIMUS and provides examples of simulated acoustic fields and ultrasound images. The accompanying article (part II) is devoted to the comparison of SIMUS with several software: Field II, k-Wave, FOCUS, and the Verasonics simulator. The MATLAB open codes for the simulator SIMUS are distributed under the terms of the GNU Lesser General Public License, and can be downloaded from https://www.biomecardio.com/MUST.

*Conclusions* – The simulations described in the accompanying paper (Part II) show that SIMUS can be used for realistic simulations in medical ultrasound imaging.

## I. Introduction

COMPUTATIONAL ultrasound imaging, which uses numerical analysis to solve problems that involve ultrasound wave propagations, has become a standard methodology in the medical ultrasound imaging community. Before considering *in vitro* or *in vivo* investigations, computational ultrasound imaging can be used, for example, to 1) analyze ultrasound sequences and arrays [1], 2) develop or optimize beamforming or post-processing algorithms [2], 3) explore multiple configurations through serial tests [3], 4) compare with peers in international challenges [4]. Among the freely available ultrasound simulators, Field II [5], [6], and k-Wave [7], [8] are arguably the most popular. These MATLAB toolboxes have widely promoted the use of ultrasound simulations for research purposes, and the number of works that use these tools has been increasing over the years (Fig. 1). There is a whole range of software packages dedicated to ultrasound imaging, available for free, as open-source or not. A non-exhaustive list of ultrasound-imaging programs is available on the k-Wave website[1]. These software programs each have their specificities, both in their application and in the numerical method: propagation to simulate acoustic pressure fields [9], [10] and/or backpropagation to also generate ultrasound images [5], [7]; two- and/or three-dimensional; solved in the time [11] or frequency [12] domain; linear [13] and/or non-linear [14], [15]; grid-based [7] or mesh-free [5]; media with homogeneous [6] or inhomogeneous speed of sound [7], [11]; or convolutional methods to quickly generate synthetic B-mode images [16], [17].

In this article, we propose a frequency-based ultrasound simulator called SIMUS. The goal was not to bring innovation in theoretical acoustics. The novelty of SIMUS lies in the computational model, which combines linear models described in several articles [18]–[24] and in Schmerr's book [25]. As detailed in the following sections, this simulator is based on far-field (Fraunhofer) and paraxial (Fresnel) acoustic equations. As we will see, the transducer elements are partitioned along the azimuth *X*-direction to enable the use of far-field equations. Roughly speaking, the paraxial approximation is valid if one does not deviate too much from the *X-Z* azimuth plane, i.e. if the angles relative to this plane are small.

SIMUS is the name of the MATLAB main function that simulates ultrasound radiofrequency (RF) signals. The conditions and assumptions under which SIMUS works are explained in this document. SIMUS is an open-source code that can be adapted by an advanced user for her/his own purpose. We created SIMUS primarily for educational and practical purposes. It was first intended for students and researchers, as they needed fast, open-source programs for their research projects [1], [26]. The ultrasound simulator SIMUS is part of the MUST toolbox

---

D. G. is with INSERM at CREATIS (Centre de Recherche en Acquisition et Traitement de l'Image pour la Santé), CNRS UMR 5220 – INSERM U1206 – Université Lyon 1 – INSA Lyon - Université Jean Monnet Saint-Etienne (e-mail: damien.garcia@inserm.fr; garcia.damien@gmail.com).

D. G. began this work when he was head of the RUBIC (Research Unit of Biomechanics and Imaging in Cardiology) at the LBUM, CRCHUM (Centre de recherche du centre hospitalier de l'Université de Montréal), Montreal, QC, Canada.

[1] http://www.k-wave.org/acousticsoftware.php

(MATLAB UltraSound Toolbox) [27], which we distribute online under the terms of the GNU Lesser General Public License v3.0 (www.biomecardio.com/MUST/). The MUST toolbox is intended for students and researchers, both novice or advanced, for teaching or research in ultrasound medical imaging. The website includes many practical examples that allow a quick understanding of the essentials of ultrasound imaging.

As with Field II [5], FOCUS [9], and k-Wave [7], MATLAB was chosen as the programming language because of its widespread use in universities and research labs, and its rich repertoire of built-in functions for data analysis, data processing, and image display. At the time of submission of this paper, only 1-D probes, rectilinear or convex, with elevation focusing, are considered in SIMUS. Although there is a growing interest in ultrasound imaging with a high number of transducer elements (e.g. 1024) and 2-D matrix arrays [28], [29], it appears that the 1-D configuration with a limited number of channels (typically 64 to 192) remains by far the most common configuration at present. The main assumptions on which SIMUS relies are

1. Linearity,
2. Scatterers acting as monopole sources,
3. Weak (single) scattering.

In essence, these hypotheses are similar to those in Field II. SIMUS, however, works in the frequency domain using a time-harmonic analysis. The frequency domain can indeed be more appropriate when dealing with bandpass signals. It can also easily consider physical aspects that depend on the frequency, such as the directivity of the elements, attenuation in the tissues, or Rayleigh scattering, for example. For the sake of clarity, the syntax of SIMUS has been standardized with that of the functions included in the MUST toolbox, and the default settings are those commonly used in medical ultrasound. SIMUS is the main program that uses another MATLAB function called PFIELD, which calculates one-way or two-way acoustic pressure fields. SIMUS and PFIELD can be used independently of the other MUST functions. SIMUS calculates the radiofrequency from the acoustic spectra generated by PFIELD. One or the other function will be mentioned depending on the context.

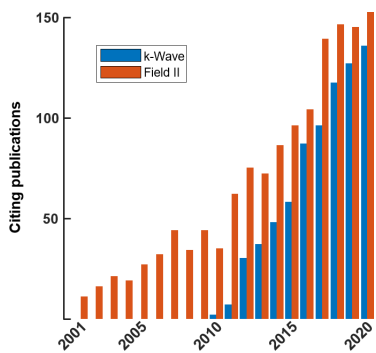

Fig. 1. Number of yearly publications that cite [5] (Field II) or [7] (k-Wave). The citation reports are from Web of Science.

This article (*Part I: theory*) is accompanied by a second one (*Part II: comparison with FieldII, k-Wave, FOCUS, and Verasonics*, Cigier A., Varray F. and Garcia D.) This first part describes the linear acoustic theory that underlies PFIELD. Several approximations and linearizations have been used. It is essential to review them to identify the limits of PFIELD and under which conditions it can be used. Part I is illustrated with simulations of ultrasound pressure fields and ultrasound images. The second article (Part II) is devoted to the comparison of the acoustic pressures generated by PFIELD with those obtained by Field II, k-Wave, FOCUS, and the Verasonics simulator.

In this first-part article, we will outline the theoretical reasoning leading to the equations included in PFIELD. We will first explain how ultrasound pressure fields are simulated and then address the generation of backscattered pressure signals. The last section will be illustrated with some realistic examples related to medical imaging.

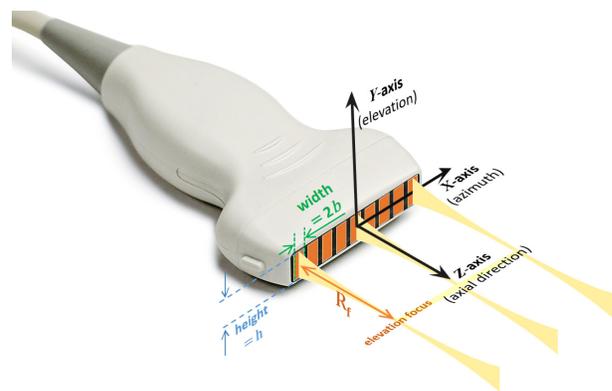

Fig. 2. Coordinate system for a rectilinear array. In this paper, the height of an element is noted $h$, and its width is $2b$. $R_f$ stands for the distance to the elevation focus.

## II. PFIELD INSIDE OUT

This section describes the theory inside PFIELD. PFIELD simulates the pressure fields in the frequency domain, i.e. it is assumed that the pressure waves have a harmonic time dependence such that the pressure is written as:

$$p(\boldsymbol{X},t) = \text{Re}\left\{\int_{-\infty}^{+\infty} P(\boldsymbol{X},\omega,t)e^{-i\omega t}\mathrm{d}\omega\right\}. \quad (1)$$

$\boldsymbol{X}$ represents a point in the radiated region of interest, $i = \sqrt{-1}$, $t$ is time, $\omega$ is the angular frequency. In the following subsections, we will describe how the pressure component $P(\boldsymbol{X},\omega,t)$ generated from one array element can be approximated, from which the backscattered echoes will be deduced. Although PFIELD also works for curvilinear arrays, the following sections describe the theory in the context of a rectilinear probe (Fig. 2). The interested reader can refer to the PFIELD code[2] to learn about the slight modifications required for a convex probe. To estimate the waves that are backscattered by a medium of point-like scatterers, one must first calculate the acoustic pressure radiated by a single element transducer and an ultrasound array. As we will see, the width ($= 2b$, see Fig. 2) and height ($h$) of the element transducers are two key parameters. Recalling that only 1-D arrays will be addressed, elevation focusing will be taken into consideration.

*A. Overview of the problem*

We will use the conventional coordinate system for a rectilinear array (Fig. 2), i.e. $X$ along the azimuthal direction, $Y$ for the elevation, and $Z$ describing the axial position. A similar system, noted in lowercase letters ($x, y, z$), will be applied for individual elements or sub-elements (Fig. 3): i.e. $x, y, z$ all equal zero at the center of an individual (sub-)element. The model starts with the Rayleigh-Sommerfeld integral, which describes an isolated element behaving like a baffled piston that vibrates on the $x$-$y$ plane (Fig. 4). From the pressure waves radiated by a single element will follow those of the ultrasound probe. The distance between a point $\mathbf{x}' = (x', y', 0)$ of the piston and a point $\mathbf{x} = (x, y, z)$ in the field (Fig. 3), is noted

$$r' = \sqrt{(x-x')^2 + (y-y')^2 + z^2}. \tag{2}$$

The distance between the center of the piston and a point $\mathbf{x}$ in the field (Fig. 3), is noted

$$r = \sqrt{x^2 + y^2 + z^2}. \tag{3}$$

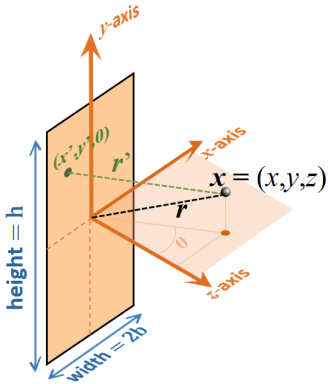

Fig. 3. Coordinate system for an individual element. The distance $r'$ is approximated by the expression (6).

Medical ultrasound one-dimensional linear and convex arrays contain an acoustic lens that focuses the ultrasound waves on the elevation plane. In SIMUS and PFIELD, the incident waves can be focused on the elevation plane at a given distance $R_f$ (Fig. 2 and Fig. 4), whose value is generally provided by the probe manufacturer. A strategy for simulating elevation focusing is to use a large parabolic element [30] or to position small elements onto a parabolic surface [31]. Another strategy is to modify the piston velocity delays along the elevation direction (Fig. 4), as described by Eq. (3.27) in [25]:

$$v(y', \omega) = v_0(\omega) e^{-ik\frac{y'^2}{2R_f}} \text{ if } |y'| \leq \frac{h}{2},\ 0 \text{ otherwise,} \tag{4}$$

with $v_0(\omega)$ being the velocity amplitude. The time delays in Eq. (4) ($= y'^2/(2cR_f)$) were derived by assuming that $y' \ll R_f$ (paraxial approximation, see Eq. (3.24) in [25]).

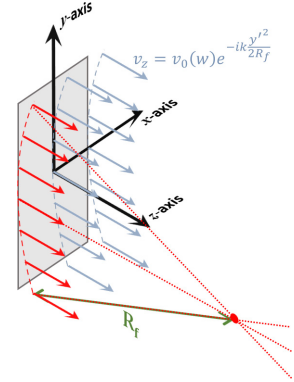

Fig. 4. An individual element acts as a baffled piston vibrating in the $z$-direction. The velocities are delayed to simulate the elevation focusing induced by an acoustic lens.

*B. Acoustic field of a single array element*

Let us consider a planar piston embedded in an infinite rigid baffle, and vibrating along the $z$-direction. The resulting harmonic pressure $P$ at position $\mathbf{x} = (x, y, z)$ is given by the Rayleigh-Sommerfeld integral (see e.g. Eq. 1 in [24] or Eq. 6.19 in [25])

$$P(\mathbf{x}, \omega, t) = \frac{k\rho c}{2i\pi} e^{-i\omega t} \int_{-b}^{b} \int_{-h/2}^{h/2} v(y', \omega) \frac{e^{ikr'}}{r'} dx' dy', \tag{5}$$

where $\rho$ is the medium density, $c$ is the speed of sound, and $k = \omega/c$ is the wavenumber. Assuming a paraxial propagation with respect to the $z$-direction, the distance $r'$ [Eq. (3)] can be rewritten, in the far field, as (see Appendix)

$$r' \approx r - x' \sin\theta + \frac{(y-y')^2}{2r}, \tag{6}$$

where the angle $\theta$ is defined in Fig. 3. The far-field pressure can thus be approximated by inserting (6) into (5) and by substituting $r$ for $r'$ in the denominator:

---

[2] https://www.biomecardio.com/MUST

$P(\mathbf{x},\omega,t)$
$$= \frac{k\rho c}{2i\pi} e^{-i\omega t} \iint_{-b,-h/2}^{b,h/2} v(y',\omega) \frac{e^{ik\left(r-x'\sin\theta + \frac{(y-y')^2}{2r}\right)}}{r} dx'dy', \quad (7)$$

which gives

$$P(\mathbf{x},\omega,t) = \frac{k\rho c}{2i\pi} \frac{e^{ikr}}{r} e^{-i\omega t} \times \ldots$$
$$\left\{\int_{-b}^{b} e^{-ikx'\sin\theta} dx'\right\}\left\{\int_{-h/2}^{h/2} v(y',\omega) e^{ik\frac{(y-y')^2}{2r}} dy'\right\}. \quad (8)$$

Including the expression of the piston velocity [Eq. (4)] in the integrand of the second integral of Eq. (8) yields

$$P(\mathbf{x},\omega,t) = \frac{k\rho c}{2i\pi} \frac{e^{ikr}}{r} v_0(\omega) e^{-i\omega t} \times \ldots$$
$$\left\{\int_{-b}^{b} e^{-ikx'\sin\theta} dx'\right\}\left\{\int_{-h/2}^{h/2} e^{-ik\frac{y'^2}{2R_f}} e^{ik\frac{(y-y')^2}{2r}} dy'\right\}. \quad (9)$$

The two integrals in the curly brackets will now be determined. The first integral in Eq. (9) yields[3]

$$\int_{-b}^{b} e^{-ikx'\sin\theta} dx' = 2b\,\text{sinc}(kb\sin\theta). \quad (10)$$

The second integral of Eq. (9) could be explicitly expressed[4] by using the imaginary error function (erfi). However, the numerical estimation of the erfi function, e.g. through estimating the Faddeeva function [32], is computationally expensive. We thus opted for the use of a Gaussian superposition model [23]. The second integral of Eq. (9) is rewritten as

$$\int_{-h/2}^{h/2} e^{-ik\frac{y'^2}{2R_f}} e^{ik\frac{(y-y')^2}{2r}} dy'$$
$$= \int_{-\infty}^{+\infty} \Pi\left(\frac{y'}{h}\right) e^{-ik\frac{y'^2}{2R_f}} e^{ik\frac{(y-y')^2}{2r}} dy'. \quad (11)$$

where $\Pi$ stands for the rectangle function. In the Gaussian superposition model, the rectangle function is approximated by a sum of Gaussians with complex coefficients (Fig. 5):

$$\Pi\left(\frac{y'}{h}\right) \approx \sum_{g=1}^{G} A_g e^{-B_g\left(\frac{y'}{h}\right)^2}. \quad (12)$$

Equation (11) thus becomes

$$\int_{-\infty}^{+\infty} \Pi\left(\frac{y'}{h}\right) e^{-ik\frac{y'^2}{2R_f}} e^{ik\frac{(y-y')^2}{2r}} dy'$$
$$\approx \sum_{g=1}^{G} A_g \int_{-\infty}^{+\infty} e^{-\alpha_g y'^2 + \beta y' + \gamma} dy', \quad (13)$$

where

$$\alpha_g = \frac{B_g}{h^2} + \frac{ik}{2}\left(\frac{1}{R_f} - \frac{1}{r}\right); \ \beta = \frac{-iky}{r}; \ \gamma = \frac{iky^2}{2r}. \quad (14)$$

Solving the right-hand side Gaussian integral[5] in Eq. (13) yields

$$\int_{-\infty}^{+\infty} \Pi\left(\frac{y'}{h}\right) e^{-ik\frac{y'^2}{2R_f}} e^{ik\frac{(y-y')^2}{2r}} dy' \approx \sum_{g=1}^{G} A_g \sqrt{\frac{\pi}{\alpha_g}} e^{\frac{\beta^2}{4\alpha_g}+\gamma}. \quad (15)$$

From Eq. (15), the second integral of Eq. (9) thus reduces to

$$\int_{-h/2}^{h/2} e^{-ik\frac{y'^2}{2R_f}} e^{ik\frac{(y-y')^2}{2r}} dy' \approx \sum_{g=1}^{G} A_g \sqrt{\frac{\pi}{\alpha_g}} e^{\frac{\beta^2}{4\alpha_g}+\gamma}. \quad (16)$$

Replacing the two integrals in (9) by their respective expressions (10) and (16) provides an estimate of the acoustic pressure generated by a single element:

$P(\mathbf{x},\omega,t) \approx$
$$\underbrace{\left\{\frac{kb}{i\pi}\rho c v_0(\omega)\right\}}_{P_{\text{Tx}}(\omega)} \frac{e^{ikr}}{r} e^{-i\omega t} \underbrace{\text{sinc}(kb\sin\theta)}_{D(\theta,k)} \underbrace{\left\{\sum_{g=1}^{G} A_g \sqrt{\frac{\pi}{\alpha_g}} e^{\frac{\beta^2}{4\alpha_g}+\gamma}\right\}}_{\delta(y,r,k)}. \quad (17)$$

The coefficients $A_g$ and $B_g$ were determined by minimizing the $\ell_2$ norm of the difference (rectangle − sum of Gaussians, Fig. 5) in the interval [−2,2]. To ensure that the sum of the complex Gaussians is real, the coefficients are the complex conjugates of the others. By default, PFIELD uses four coefficients:

$$\begin{aligned}
&A_1 = 0.187 + 0.275i, B_1 = 4.558 - 25.59i,\\
&A_2 = 0.288 - 1.954i, B_2 = 8.598 - 7.924i,\\
&A_3 = \overline{A_1}, B_3 = \overline{B_1},\\
&A_4 = \overline{A_2}, B_4 = \overline{B_2}.
\end{aligned} \quad (18)$$

More coefficients can be used to obtain more numerical accuracy (at the expense of computing time). Lists with up to 25 coefficients are provided in [23], [33]. It should be noted, however, that obtaining realistic ultrasound images does not require very fine numerical precision. By way of example, Fig. 6 shows the effects of the number of Gaussians on focused pressure fields generated by a cardiac phased array. Four Gaussians led to small differences compared with ten Gaussians.

---

[3] https://www.wolframalpha.com/input/?i=int+exp%28-i*k*S*x%29+from+x+%3D+-b+to+b

[4] https://www.wolframalpha.com/input/?i=int+exp%28-i*%28P*Y%5E2%2BQ*Y%2BR%29%29+from+Y+%3D+-h%2F2+to+h%2F2

[5] https://mathworld.wolfram.com/GaussianIntegral.html

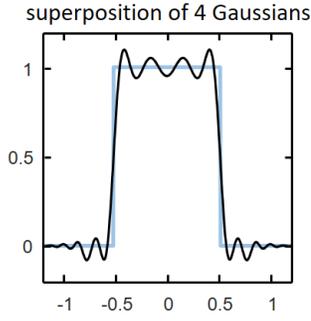

Fig. 5. The rectangle function can be approximated by the sum of Gaussians, which simplifies the estimation of the integral in Eq. (11).

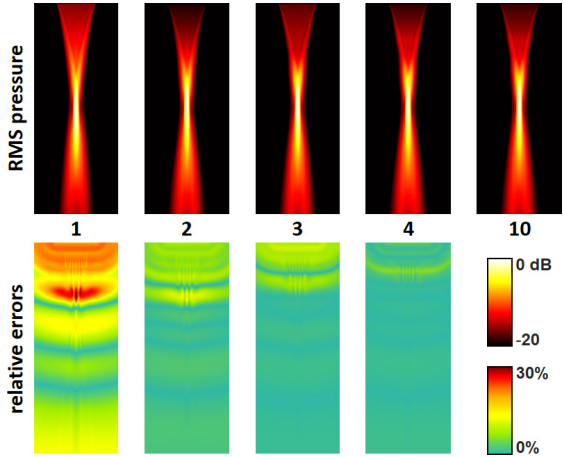

Fig. 6. Effects of the number of Gaussians in the Gaussian superposition model. Top row: focused pressure fields simulated with PFIELD for a 64-element 2.7-MHz cardiac phased array (see Fig. 9). The middle line indicates the number of Gaussians. Bottom row: relative errors with respect to the pressure field obtained with 10 Gaussians.

In practice, it is not the velocity of the element that is known, but the acoustic pressure generated by an element, measured for example with a hydrophone. The first term in brackets (dimensionally homogeneous to pressure) in the expression (17) represents the spectrum of the transmit pressure pulse (up to a constant multiplier), noted $P_{Tx}(\omega)$. The sine cardinal sinc term represents the $x$-directivity of one element, noted $D(\theta, k)$. The last term in brackets is related to the elevation focusing. It is homogeneous to a distance and is noted $\delta(y, r, k)$. Using these notations, the acoustic pressure of one element finally reduces to

$$P(x, \omega, t) \approx P_{Tx}(\omega) \frac{e^{ikr}}{r} D(\theta, k) \delta(y, r, k) e^{-i\omega t}. \quad (19)$$

More generally, if the transmission is delayed by $\Delta\tau$, the wave field is given by

$$P(x, \omega, t) \approx P_{Tx}(\omega) \frac{e^{ikr}}{r} D(\theta, k) \delta(y, r, k) e^{i\omega\Delta\tau} e^{-i\omega t}. \quad (20)$$

It is recalled that the position variables $(x, \theta, r)$ in (20) are relative to the center of the element (Fig. 3).

### C. Acoustic field of a rectilinear array

The expression (20) models acoustic waves *radiated by one element*. To derive this expression, the distance $r$ with respect to the element [see Eq. (6)] was simplified by using a Fraunhofer (far-field) approximation in the azimuthal $x$-direction, and a Fresnel (paraxial) approximation in the elevation $y$-direction. In order to fulfill the far-field condition, it may be necessary to split the array elements into $\nu$ sub-elements, in the azimuthal $x$-direction, if they are too wide (Fig. 7). Radiation patterns that result from splitting are given in Appendix C for element transducers of different widths. In PFIELD, $\nu$ is chosen so that a sub-element width $(=2b/\nu)$ is not greater than the minimal wavelength (defined at -6dB):

$$\nu = \left\lceil \frac{2b}{\lambda_{\min}} \right\rceil, \quad (21)$$

where $\lceil\ \rceil$ is the ceiling function. As an indication, the number $\nu$ of simulated sub-element(s) per array element is typically 1 for cardiac phased arrays, and 2 for linear arrays. If an array contains $N$ elements, the total number of sub-elements is thus $(\nu N)$. Given the properties of linearity, the acoustic wavefield produced by an $N$-element array can be modeled by superimposing the $(\nu N)$ individual sub-element models described by (20):

$$\boxed{\begin{aligned} P(X, \omega, t) \approx \\ P_{Tx}(\omega) e^{-i\omega t} \sum_{n=1}^{\nu N} W_n \frac{e^{ikr_n}}{r_n} D(\theta_n, k) \delta(y_n, r_n, k) \, e^{i\omega\Delta\tau_n}. \end{aligned}} \quad (22)$$

In (22), the position variables $(\theta_n, y_n, r_n)$ are relative to the center of the $n^{\text{th}}$ sub-element. The position $X = (X, Y, Z)$ is relative to the coordinate system of the array depicted in Fig. 2. If $X_{c,n}$ stands for the abscissa of the $n^{\text{th}}$ sub-element centroid, then

$$r_n = \sqrt{(X - X_{c,n})^2 + Y^2 + Z^2}, \quad (23)$$

and $\sin\theta_n = (X - X_{c,n})/\sqrt{(X - X_{c,n})^2 + Z^2}$.

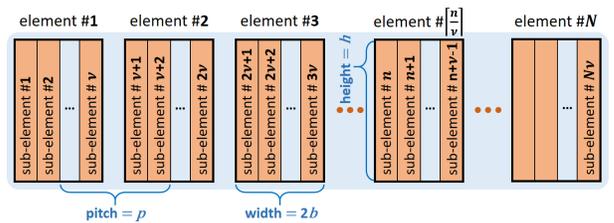

Fig. 7. To ensure that the far-field condition is met, the $N$ array elements are each divided into $\nu$ sub-elements.

We here used $Z_{c,n} = 0$ since we only address rectilinear arrays in this paper. Let the leftmost to rightmost sub-elements be numbered sequentially, from 1 to $n$. In the case of a uniform linear array of pitch $p$ (Fig. 7), it can be shown that its centroid abscissa can be written as

$$X_{c,n} = \frac{p}{2}\left(2\left\lceil \frac{n}{v} \right\rceil - N - 1\right) + \frac{b}{v}(2(n-1)(\text{mod }v) - v + 1) \quad (24)$$

The term $\lceil \frac{n}{v} \rceil$ corresponds to the number of the element to which the sub-element $n$ belongs. Note that $X_{c,n}$ (and $Z_{c,n} \neq 0$) must be modified for a convex array (see the PFIELD code for details). The transmit delay $\Delta \tau_n = \Delta \tau_{\lceil n/v \rceil}$ in (22) is that of the $\lceil n/v \rceil^{\text{th}}$ element. Each element has been weighted by $W_n = W_{\lceil n/v \rceil}$ to consider transmit apodization. The equation (22) is the backbone of PFIELD. Once the transmit pressure $P_{\text{Tx}}(\omega)$ is given in the frequency domain, it allows simulation of realistic fields of acoustic pressure produced by an ultrasound array. In PFIELD, the transmit pressure is generated by convolving the one-way response of the transducer with a windowed sine, as explained in the next section. An advanced user can define her own $P_{\text{Tx}}(\omega)$ spectrum, derived from experimental measurements for example. From the acoustic reciprocity principle, Eq. (22) can also be applied to derive the backscattered echoes, as will be explained in section IV.

## III. SPECTRUM OF THE TRANSMIT PRESSURE

The general expression (22) contains the spectrum of the transmit pressure $P_{\text{Tx}}(\omega)$. An advanced user can easily include the spectrum of his/her choice in the code. In SIMUS and PFIELD, the default transmit pressure waveform is obtained by convolving a rectangularly-windowed sinusoid (a "perfect" pulse) with the point spread function (PSF) of the transducer. The angular frequency of the sinusoid is that of the transducer ($\omega_c = 2\pi f_c$, the center frequency $f_c$ being provided by the manufacturer). The temporal width $T$ of the rectangular window is defined in terms of the number of wavelengths $n_\lambda$ by $T = n_\lambda / f_c = 2\pi n_\lambda / \omega_c$. The spectrum of the rectangularly-windowed pulse is given by

$$S_{\text{P}}(\omega) = i\left[\text{sinc}\left(T\frac{\omega - \omega_c}{2}\right) - \text{sinc}\left(T\frac{\omega + \omega_c}{2}\right)\right] \quad (25)$$

The spectrum of the transducer PSF is defined by a generalized Gaussian window that depends on two positive parameters $p$ and $\sigma$:

$$S_{\text{T}}(\omega) = e^{-\left(\frac{|\omega - \omega_c|}{\sigma \omega_c}\right)^p} \quad (26)$$

It is designed so that its pulse-echo response has a given bandwidth $\omega_b$ at -6 dB (Fig. 8). The pulse-echo fractional bandwidth is generally given by the manufacturer in percent. For example, a 65% bandwidth means that the frequency bandwidth of the response is such that $\omega_b = 0.65\,\omega_c$. To determine both $p$ and $\sigma$, it is postulated that $S_{\text{T}}(0) = 2^{-126}$ (the smallest positive single-precision floating-point number; PFIELD and SIMUS work in single precision). It follows that (see Appendix)

$$S_{\text{T}}(\omega) = e^{-\ln 2\left(\frac{2|\omega - \omega_c|}{\omega_b}\right)^p}, \text{ with } p = \ln 126 / \ln\left(2\frac{\omega_c}{\omega_b}\right). \quad (27)$$

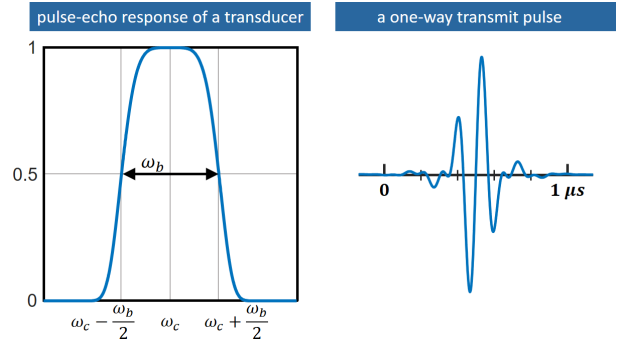

Fig. 8. *Left* – Pulse-echo response of the transducer. $\omega_c$ is the center angular frequency ($\omega_c = 2\pi f_c$). $\omega_b$ is the angular frequency bandwidth. *Right* – Example of a one-way transmit pulse (center frequency = 7.6 MHz, fractional bandwidth = 77%, excitation pulse = 1 wavelength).

Fig. 8 (left panel) depicts an example of the simulated transducer response for a pulse-echo fractional bandwidth of 77%. PFIELD does not include the electrical-acoustic conversions that occur in the piezoelectric elements. The units of the simulated acoustic and electrical signals are thus arbitrary (not Pa or V). By using this simplified representation and writing the convolution in the frequency domain, the spectrum of the transmit pressure $P_{\text{Tx}}(\omega)$ is given by this proportionality relationship:

$$P_{\text{Tx}}(\omega) \propto S_{\text{P}}(\omega)\sqrt{S_{\text{T}}(\omega)}. \quad (28)$$

A square root is needed since the transducer response $S_{\text{T}}(\omega)$ is two-way (transmit + receive). Fig. 8 (right panel) presents a transmit pulse (one-way) in the temporal domain.

## IV. SIMUS INSIDE OUT

### A. Echoes received by a sub-element

SIMUS uses PFIELD and point-like scatterers to simulate radiofrequency (RF) ultrasound signals. These scatterers become individual monopole point sources when an incident wave reaches them. They do not acoustically interact with each other according to the assumption of single weak scattering. Each scatterer is defined by its reflection coefficient ($\mathcal{R}_s$), which describes how much amplitude of a wave is reflected. Although some tissues, such as blood, are governed by Rayleigh scattering [34], the $\mathcal{R}_s$ coefficients are assumed constant, i.e. independent of frequency and incidence angle. From (22), the pressure signal received by a scatterer #$s$ located at $X_s = (X_s, Y_s, Z_s)$ is

$$P(X_s, \omega, t) \approx$$
$$P_{\text{Tx}}(\omega)e^{-i\omega t}\sum_{n=1}^{vN} W_n \frac{e^{ikr_{ns}}}{r_{ns}} D(\theta_{ns}, k)\delta(Y_s, r_{ns}, k)\, e^{i\omega\Delta\tau_n}, \quad (29)$$

where $r_{ns} = \sqrt{(X_s - X_{c,n})^2 + Y_s^2 + Z_s^2}$ and $\sin\theta_{ns} = (X_s - X_{c,n})/\sqrt{(X_s - X_{c,n})^2 + Z_s^2}$. The principle of acoustic reciprocity dictates that an acoustic response remains identical when the source and receiver are exchanged. Expression (19) can therefore give the pressure received by the $m^{\text{th}}$ sub-element from a scatterer $s$, after accounting for its reflection coefficient $\mathcal{R}_s$:

$$P^{\text{se}}{}_{ms}(\omega,t) \approx \mathcal{R}_s \underbrace{P(\boldsymbol{X}_s,\omega,t)}_{P(\boldsymbol{X}_s,\omega)e^{-i\omega t}} \frac{e^{ikr_{ms}}}{r_{ms}} D(\theta_{ms},k)\delta(Y_s,r_{ms},k), \quad (30)$$

where $r_{ms} = \sqrt{(X_s - X_{c,m})^2 + Y_s^2 + Z_s^2}$ and $\sin\theta_{ms} = (X_s - X_{c,m})/\sqrt{(X_s - X_{c,m})^2 + Z_s^2}$. In (30), the superscript "se" means "sub-element". Inserting (29) in (30), it follows that:

$$\begin{aligned}P^{\text{se}}{}_{ms}(\omega,t) &\approx \mathcal{R}_s P_{\text{Tx}}(\omega) e^{-i\omega t} \\ &\times \left[\sum_{n=1}^{vN} W_n \frac{e^{ikr_{ns}}}{r_{ns}} D(\theta_{ns},k)\delta(Y_s,r_{ns},k)\, e^{i\omega\Delta\tau_n}\right] \\ &\times \frac{e^{ikr_{ms}}}{r_{ms}} D(\theta_{ms},k)\delta(Y_s,r_{ms},k).\end{aligned} \quad (31)$$

The expression (31) is the acoustic pressure backscattered by a single scatterer and received by the $m^{\text{th}}$ sub-element. Assuming now that there is a total of $S$ scatterers, the combination of their independent effect (single scattering assumption) gives the total pressure received by the $m^{\text{th}}$ sub-element:

$$\begin{aligned}P^{\text{se}}{}_m(\omega,t) &\approx P_{\text{Tx}}(\omega) e^{-i\omega t} \sum_{s=1}^{S}\Big\{\mathcal{R}_s \\ &\times \left[\sum_{n=1}^{vN} W_n \frac{e^{ikr_{ns}}}{r_{ns}} D(\theta_{ns},k)\delta(Y_s,r_{ns},k)\, e^{i\omega\Delta\tau_n}\right] \\ &\times \frac{e^{ikr_{ms}}}{r_{ms}} D(\theta_{ms},k)\delta(Y_s,r_{ms},k)\Big\}.\end{aligned} \quad (32)$$

The expression (32) is for an individual sub-element $\#m$. The pressure wave $P^e(\omega,t)$ received by one transducer element is the coherent sum of the pressures received by all its sub-elements (Fig. 7). For the element $\#\lceil\frac{n}{v}\rceil$:

$$P^e{}_{\lceil\frac{n}{v}\rceil}(\omega,t) = \sum_{\mu=0}^{v-1} P^{\text{se}}{}_{n+\mu}(\omega,t) \quad (33)$$

The theoretical pressures were all derived for a single angular frequency $\omega = 2\pi f$. The full-band waveforms can be obtained by summation in the frequency domain through an inverse fast Fourier transform.

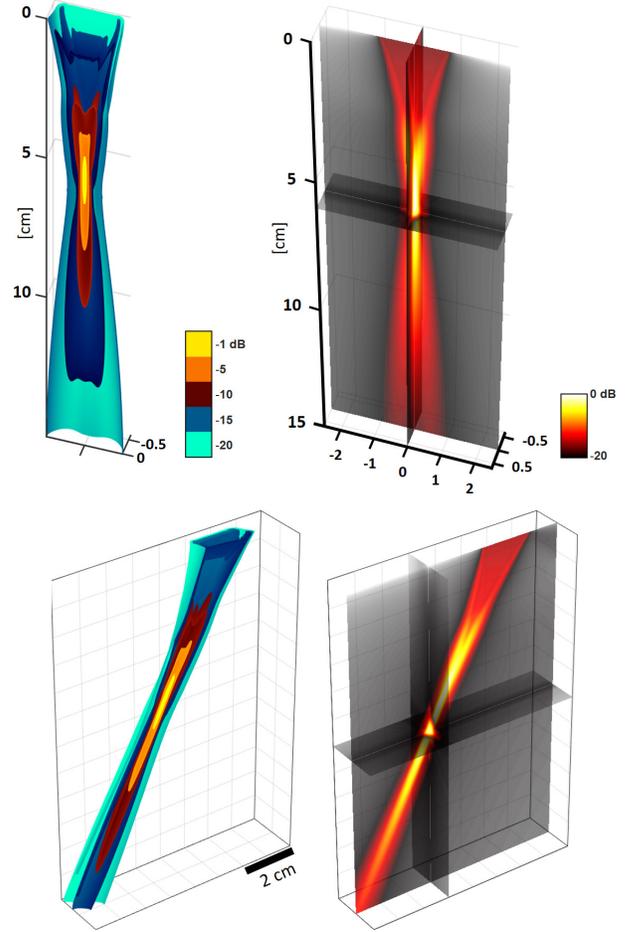

Fig. 9. Focused pressure fields simulated with PFIELD for a 64-element 2.7-MHz P4-2v cardiac phased array (kerf width = 50 µm, pitch = 0.3 mm, fractional bandwidth = 74%, elevation focus = 6 cm). These RMS (root mean square) acoustic fields illustrate emission sequences such as those used in standard transthoracic echocardiography, with focusing in the axial direction.

### B. Radiofrequency signals

The spectrum of the radiofrequency RF signal of the $m^{\text{th}}$ sub-element is related to the received acoustic pressure (32) by

$$\text{RF}^{\text{se}}{}_m(\omega,t) \propto \sqrt{\mathcal{S}_{\text{T}}(\omega)}\, P^{\text{se}}{}_m(\omega,t), \quad (34)$$

where it is recalled that $\sqrt{\mathcal{S}_{\text{T}}(\omega)}$ is the one-way transducer response. Alike (33), the RF signal related to one element is the coherent sum of the RF signals of its sub-elements.

## V. THE 2-D CASE

PFIELD and SIMUS can also work in two dimensions to speed up calculations, for example, when rapid testing is required. In a two-dimensional (2-D) $x$-$z$ domain, the piston-like element generates a normal velocity that is constant everywhere (i.e. in $[-\infty, +\infty]$) in the $y$-direction. This situation is obtained when $h$ (element height) and $R_f$ (distance to elevation focus) both tend to $+\infty$. In this limit case, the rectangular function in Eq. (12) becomes 1 for any $y'$, and the coefficients reduce to $G = 1$, with $A_1 = 1$ and $B_1 = 0$. Under these conditions, $\beta^2/(4\alpha_1) + \gamma = 0$ and $\sqrt{(\pi/\alpha_1)} = \sqrt{(2i\pi r/k)}$ [see Eq. (14) and (15)]. In 2-D, from (17), the acoustic pressure generated by a single element thus reads

$$P_{\text{2-D}}(\mathbf{x}, \omega, t) \approx \left\{\frac{kb}{i\pi}\rho c v_0(\omega)\right\}\frac{e^{ikr}}{r}e^{-i\omega t}\,\text{sinc}(kb\sin\theta)\sqrt{\frac{2i\pi r}{k}}, \quad (35)$$

which can be rewritten as

$$P_{\text{2-D}}(\mathbf{x}, \omega, t) \approx \sqrt{2b}\underbrace{\left\{\sqrt{\frac{kb}{i\pi}}\rho c v_0(\omega)\right\}}_{P_{\text{Tx}}(\omega)}\frac{e^{ikr}}{\sqrt{r}}e^{-i\omega t}\underbrace{\text{sinc}(kb\sin\theta)}_{D(\theta,k)}. \quad (36)$$

The acoustic wave (22) radiated by an $N$-element array then becomes

$$P_{\text{2-D}}(\mathbf{X}, \omega, t) \approx \sqrt{2b}\,P_{\text{Tx}}(\omega)e^{-i\omega t}\sum_{n=1}^{\nu N}W_n\frac{e^{ikr_n}}{\sqrt{r_n}}D(\theta_n, k)\,e^{i\omega\Delta\tau_n}. \quad (37)$$

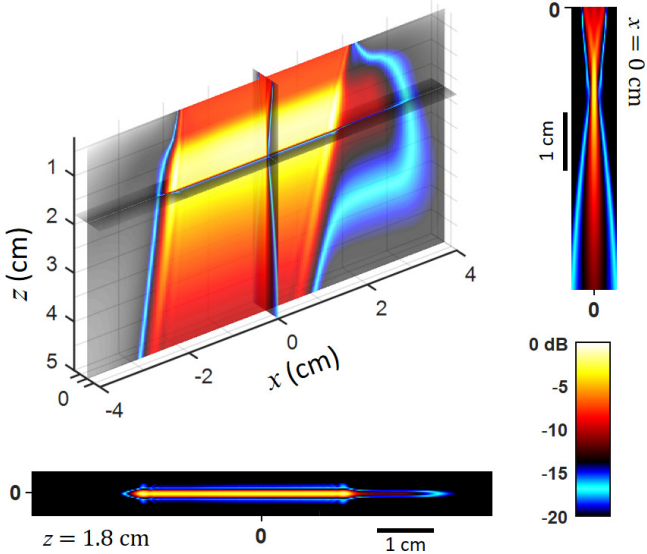

Fig. 10. Plane-wave pressure field simulated with PFIELD for a 128-element 7.6-MHz L11-5v linear array (kerf width = 30 μm, pitch = 0.27 mm, fractional bandwidth = 77%, elevation focus = 1.8 cm). This RMS (root mean square) acoustic field illustrates an emission sequence (here, tilt angle = 10°) such as that used in "ultrafast" compound plane-wave imaging, without focusing in the axial direction [35]. The two insets represent focal and elevational planes.

Similarly, the pressure received by the $m^{\text{th}}$ sub-element (32) in a 2-D space reduces to

$$P_{\text{2-D}_m}^{\text{se}}(\omega, t) \approx (2b)\,P_{\text{Tx}}(\omega)e^{-i\omega t}\sum_{s=1}^{S}\left\{\mathcal{R}_s \right.$$
$$\times \left[\sum_{n=1}^{\nu N}W_n\frac{e^{ikr_{ns}}}{\sqrt{r_{ns}}}D(\theta_{ns}, k)\,e^{i\omega\Delta\tau_n}\right] \quad (38)$$
$$\left. \times \frac{e^{ikr_{ms}}}{\sqrt{r_{ms}}}D(\theta_{ms}, k)\right\}.$$

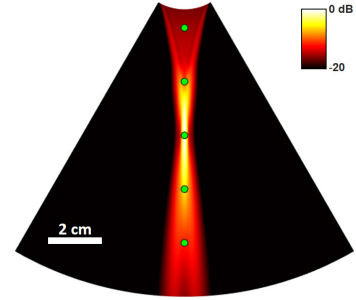

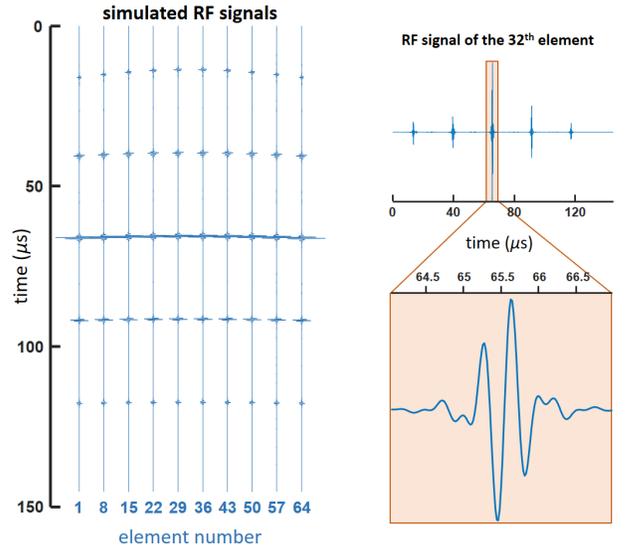

Fig. 11. RF signals simulated with SIMUS for a P4-2v phased array transmitting a focused wave in a medium that contains five scatterers.

## VI. BAFFLE IMPEDANCE AND DISPERSIVE MEDIUM

### A. Finite impedance baffle

The general expressions (22) and (32) were derived by assuming that the baffle in which the transducer element is embedded has an infinite acoustic impedance (rigid baffle). It might be recommended to use a finite baffle impedance to obtain more realistic radiation patterns [21], [22]. In such a case, an obliquity factor must be added to the element directivity $D(\theta, k)$ [see Eq. (36)]. The expressions of the obliquity factors for a nonnegative finite impedance are given in [21] (null impedance = "soft" baffle) and [22] (positive impedance). The element directivity becomes

$$\begin{aligned}\text{rigid:} \quad & D(\theta, k) = \text{sinc}(kb \sin\theta). \\ \text{soft:} \quad & D(\theta, k) = \text{sinc}(kb \sin\theta) \cos\theta. \\ \text{otherwise:} \quad & D(\theta, k) = \text{sinc}(kb \sin\theta) \frac{\cos\theta}{\cos\theta + \zeta},\end{aligned} \quad (39)$$

with $\zeta$ standing for the ratio between the medium and baffle impedances. For a baffle of impedance 2.8 MRayl (epoxy) adjacent to soft tissues of impedance 1.6 MRayl, $\zeta = 1.75$ [22]. By default, the baffle is soft in PFIELD and SIMUS.

### B. Attenuation

In addition to the distance-dependent decrease in amplitude caused by the inverse law (or inverse square-root law in 2-D), an ultrasound wave propagating in tissues is attenuated through scattering and absorption. In SIMUS, scattering is governed by the reflectivity coefficients $\mathcal{R}_s$. In the current version of the proposed ultrasound simulator, the medium is assumed non-dispersive, which means that waves of different wavelengths travel at the same phase velocity ($= c$). In PFIELD and SIMUS, absorption can be included in the amplitude by adding a frequency-dependent multiplicative loss term. For numerical reasons (a recursive multiplication is used to calculate the exponential terms), this frequency dependence is linear in the current version of PFIELD. Amplitude absorption is obtained by substituting $e^{i(k+ik_a)r}$ for $e^{ikr}$, where $(k_a)$ is given by:

$$k_a = \frac{\alpha_{\text{dB}}}{8.69} \frac{kc}{2\pi\, 10^4}. \quad (40)$$

The attenuation coefficient $\alpha_{\text{dB}}$ is in dB/cm/MHz. A typical value for soft tissues is 0.5 dB/cm/MHz [36]. The $10^4$ accounts for the unit conversion (cm·MHz to m·Hz). The scalar 8.69 is an approximation of $20/\ln(10)$ (see Eq 4.4 in [37]).

### VII. RMS PRESSURE FIELDS

Equation (22) gives the acoustic pressure field for a given angular frequency $\omega$. This is the expression used by the SIMUS code when it calls PFIELD. When used alone, the PFIELD MATLAB code returns, by default, the root-mean-square RMS pressure field. If we omit the $(e^{-i\omega t})$ term in (22) and define $P(X, \omega)$ by the relationship $P(X, \omega, t) \stackrel{\text{def}}{=} P(X, \omega)e^{-i\omega t}$, the RMS pressure field is given by

$$P_{\text{RMS}}(X) = \sqrt{\int_0^{2\omega_c} P(X, \omega)^2 \, d\omega}. \quad (41)$$

The integral can be estimated by using a midpoint Riemann sum with $N_f$ equally spaced frequency samples:

$$P_{\text{RMS}}(X) \approx \sqrt{\Delta\omega \sum_{j=0}^{N_f} P\left(X, 2\left(\frac{j}{N_f}\right)\omega_c\right)^2}. \quad (42)$$

The partition width $\Delta\omega$ must be small enough to ensure a proper approximation, but not too small to avoid computational overload with a large $N_f$. The angular frequency step is chosen so that the phase increment in (20) be smaller than $2\pi$ everywhere in the radiated region of interest (roi), i.e. $\Delta\omega$ must verify $[(\Delta\omega/c)r + \Delta\omega\Delta\tau] < 2\pi$ for any distance $r$ and transmit delay $\Delta\tau$, i.e. $\Delta\omega = \min_{\text{roi}}\{2\pi/(r/c + \Delta\tau)\}$.

Once the transmit delays $\Delta\tau_n$ and the parameters of the array are given, the equations (22) and (42) yield the RMS pressure field at the location specified by $X$. Fig. 9 and Fig. 10 illustrate two examples of acoustic fields simulated with PFIELD: focused and plane wavefronts with a cardiac and linear array, respectively. In these examples, the 3-D equation was applied to take the elevation focusing into account.

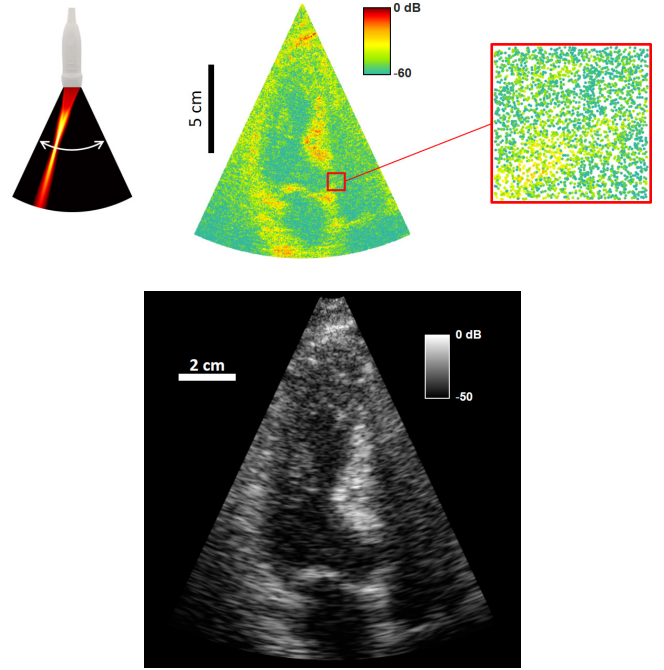

Fig. 12. Simulation of a three-chamber-view echocardiographic image. RF signals were simulated with SIMUS by using 39,500 scatterers with predefined reflection coefficients (top row). 128 focused waves were transmitted to create the 128 scanlines of the B-mode image.

### VIII. MEDICAL ULTRASOUND IMAGES WITH SIMUS

Medical ultrasound images, such as B-mode or color Doppler images, can be generated by simulating RF signals through Eq. (32). Once RF signals are obtained, they can be post-processed (e.g. using I/Q demodulation, beamforming, clutter filtering, phase analysis, envelope detection…) by using the MATLAB codes available in the MUST toolbox[6]. A series of RF signals given by SIMUS in a simplified configuration (a focused wave that radiates five scatterers) is displayed in Fig. 11. Realistic ultrasound images can be obtained when using a large number of point-like scatterers, as explained in the next paragraphs.

*A. B-mode imaging*

Fig. 12 illustrates the simulation of a long-axis echocardiographic image (three-chamber view) by scanning the left heart with a series of focused waves. The synthetic myocardial tissue contained 39,500 randomly distributed point scatterers (density = 10 scatterers per wavelength squared) whose reflection coefficients are shown in Fig. 12. The fan-shaped B-mode image consists of 128 scanlines. Each scanline was constructed from one tilted focused wave generated by a 2.7-MHz phased array (64 elements, 0.3-mm pitch, 74% fractional bandwidth, 6-cm elevation focus). RF signals were simulated with SIMUS, then I/Q demodulated and beamformed using a delay-and-sum with an optimal f-number [38]. The B-mode image was obtained by log-compressing the real envelopes.

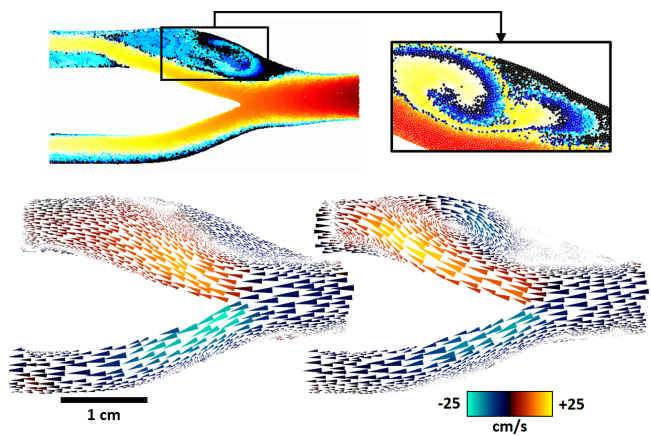

Fig. 13. Simulation of color Doppler and vector Doppler in a 2-D carotid bifurcation. The displacements of ~26,000 blood scatterers (top row) were simulated by SPH (smoothed particle hydrodynamics), a Lagrangian CFD method [26]. The inset is at another time. RF signals were simulated with SIMUS by using unsteered plane waves. Color Doppler (bottom row, red-blue patterns) and vector Doppler (bottom row, arrows) were generated by post-processing the I/Q signals (beamforming, autocorrelator, robust smoothing) with tools available in the MUST toolbox. Adapted from [26].

*B. Color flow imaging and Vector flow imaging*

SIMUS (and the tools available in the MUST toolbox [27]) can also be used to simulate realistic color flow images (color Doppler) and vector flow images. Two-dimensional simulations of color and vector Doppler in a 2-D carotid bifurcation were introduced in a previous work that combined CFD (computational fluid dynamics) by SPH (smooth particle hydrodynamics) with SIMUS [26]. In that paper, radiofrequency RF signals were simulated by SIMUS using a plane-wave-imaging sequence (non-tilted waves transmitted by a 128-element linear array). After I/Q demodulation and beamforming with two distinct subapertures (see details in [26]), a one-lag slow-time autocorrelator was applied to recover 2-D velocity fields. The two beamforming subapertures provide two Doppler images with significantly different angles, which makes it possible to derive 2-D velocity vectors. This approach has been applied successfully *in vivo* with sub-Nyquist RF sampling [39] and can be generalized with more than two subapertures [40]. Fig. 13 provides two snapshots adapted from [26].

## IX. RESULTS & DISCUSSION

Comparisons against other software packages (Field II, k-Wave, FOCUS, Verasonics), as well as the advantages and limitations of SIMUS, are discussed in the accompanying article (part II).

## X. CONCLUSION

The assumptions and simplifications included in PFIELD and SIMUS make its theory and numerical time-harmonic analysis convenient. The examples show that realistic ultrasound images can be created for educational and research purposes. How PFIELD compares to Field-II, k-Wave, FOCUS, and Verasonics is detailed in the accompanying article (part II). The current version of PFIELD (2021), although it includes the 3-D acoustic equation for elevation focusing, is limited to one-dimensional, linear, or convex ultrasonic transducers. A volume version is planned. Nevertheless, based on the far-field equations described in this paper, and keeping similar hypotheses, an advanced user could adapt SIMUS for the simulation of matrix arrays and volumetric acoustic fields.

ACKNOWLEDGMENT

I thank my former students from Montreal and Lyon who tested and re-tested the different versions and contributed to the improvement of the MUST toolbox. In particular: Jonathan Porée, Daniel Posada, Julia Faurie, Craig Madiena, Vincent Perrot. As well as my colleagues from CREATIS, Lyon, Olivier Bernard, and François Varray for making the most of MUST. This work was supported in part by the LABEX CeLyA (ANR-10-LABX-0060) of Université de Lyon, within the program "Investissements d'Avenir" (ANR-16-IDEX-0005) operated by the French National Research Agency.

---

[6] https://www.biomecardio.com/MUST

APPENDIX

*A. Paraxial and far-field distances*

Equation (2) can be rewritten as

$$r' = z\sqrt{\left[1 + \frac{(x-x')^2}{z^2}\right] + \frac{(y-y')^2}{z^2}}. \quad (43)$$

In the paraxial (Fresnel) approximation, $(y-y')^2/z^2 \ll [1 + (x-x')^2/z^2]$. One can thus write[7]

$$r' \approx \sqrt{z^2 + (x-x')^2} + \frac{(y-y')^2}{2\sqrt{z^2+(x-x')^2}}. \quad (44)$$

In the paraxial approximation, one has also $r^2 \approx x^2 + z^2$. The expression (44) can thus be approximated by

$$r' \approx r\sqrt{1 + \frac{x'^2}{r^2} - 2\frac{xx'}{r^2}} + \frac{(y-y')^2}{2r\sqrt{1+\frac{x'^2}{r^2}-2\frac{xx'}{r^2}}}. \quad (45)$$

By keeping the 1st order of $x'$ in the first square root and the 0th order in the second (far-field approximation), we obtain

$$r' \approx r\left(1 - \frac{xx'}{r^2}\right) + \frac{(y-y')^2}{2r}. \quad (46)$$

Because $\sin\theta = x/\sqrt{x^2+z^2} \approx x/r$ (Fig. 3), (46) reduces to Eq. (6).

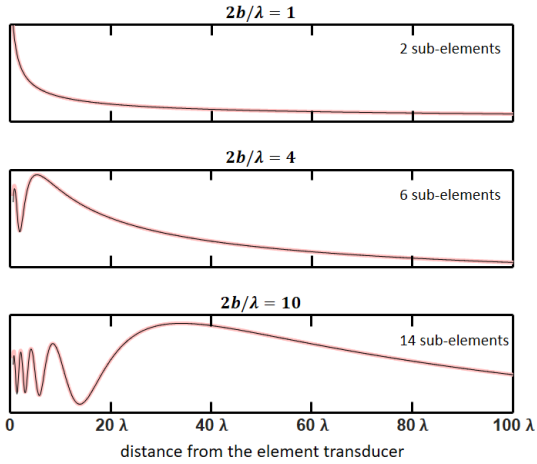

Fig. 14. Comparison of on-axis pressure magnitudes (arbitrary units) generated by one element transducer at the center frequency (2-D acoustics): element splitting + far-field (thin black lines) vs. Rayleigh-Sommerfeld integral (thick pink lines). The number of sub-elements was given by Eq. (21). The width of the element is $2b$.

*B. Spectrum of the transducer PSF*

We need a $\omega_b$ bandwidth at -6 dB. Therefore, from (26):

[7] $\sqrt{a+x} = \sqrt{a} + x/(2\sqrt{a}) + O(x^2)$

$$\mathcal{S}_t\left(\omega_c \pm \frac{\omega_b}{2}\right) = e^{-\left(\frac{\omega_b}{2\sigma\omega_c}\right)^p} = \frac{1}{2} \quad (47)$$

One can thus deduce $\sigma$:

$$\sigma = \frac{1}{\sqrt[p]{\ln 2}}\frac{\omega_b}{2\omega_c}. \quad (48)$$

By assuming that

$$\mathcal{S}_t(0) = \mathcal{S}_t(2\omega_c) = e^{-\frac{1}{\sigma^p}} = 2^{-126}, \quad (49)$$

the expressions (48) and (49) yield $p$ [Eq. (27)].

*C. Element splitting and far-field equations*

In MUST, the transducer elements are partitioned to enable the use of far-field equations (Fig. 7). The far-field approximation is valid if $2xx'/r^2 \ll 1$ [see Eq. (45) and (46)]. After partitioning, $x'$ can be as large as $b/M$ ($M$ being the number of partitions per element), and $x$ can be as large as $r$. The inequality thus becomes $(2b/M)/r \ll 1$, which means that the width of a sub-element must be sufficiently small compared to the distance to the point of interest. In PFIELD, we choose the default condition $\lambda/r \ll 1$ so that the number of partitions per element is determined by Eq. (21). Fig. 14 compares on-axis pressures (using 2-D acoustics) as calculated with element splitting [Eq. (37)] and far-field equations against those derived from the Rayleigh-Sommerfeld integral [Eq. (2.46) in [25]].

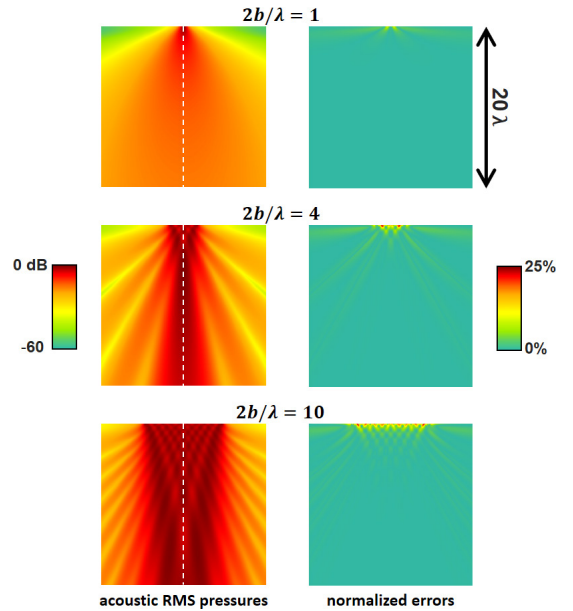

Fig. 15. SIMUS-derived radiation patterns (RMS pressures) generated by one element transducer at the center frequency with 2-D acoustics (1st column). The normalized errors (2nd column) are with respect to the Rayleigh-Sommerfeld integral. The patterns along the white dashed lines are in Fig. 14. The width of the element is $2b$.

Fig. 15 displays the radiation patterns generated by SIMUS and provides the normalized errors with respect to the Rayleigh-Sommerfeld integral. It can be seen that element splitting enables accurate pressure estimates even at small distances from the transducer. Note that the number of partitions can be optionally adjusted by the user to refine the results near the element transducer. This may however have a limited impact due to round-off errors related to the $1/r$ term in the equations. Specially adapted software such as FOCUS[8] removes this singularity, allowing high accuracy to be achieved in the near-field region [41].

---

[8] https://www.egr.msu.edu/~fultras-web

Garcia D., *SIMUS* – Part I